\documentclass[twocolumn,showpacs,preprintnumbers,pre]{revtex4}
\usepackage{graphicx} 
\usepackage{dcolumn} 
\usepackage{bm} 

\begin{document}

\title{Experimental Evidence of Fragile-to-Strong Dynamic Crossover in DNA
Hydration Water}
\author{Sow-Hsin~Chen,$^{1}$\footnote{Author to whom correspondence
should be addressed. Electronic mail: sowhsin@mit.edu}
Li~Liu,$^{1}$ Xiangqiang~Chu,$^{1}$ Yang~Zhang,$^{1}$
Emiliano~Fratini,$^{2}$ Piero~Baglioni,$^{2}$
Antonio~Faraone,$^{3}$ and Eugene~Mamontov$^3$}

\address{$^{1}$Department of Nuclear Science and Engineering,
Massachusetts Institute of Technology,\\ Cambridge MA 02139 USA
$^{2}$Department of Chemistry and CSGI, University of Florence,\\
via della Lastruccia 3, 50019 Florence, Italy $^{3}$Department of
Material Science and Engineering, University of Maryland, College
Park, MD 20742 USA and NIST Center for Neutron Research,
Gaithersburg, MD 20899-8562 USA}

\date{\today}

\begin{abstract}
We used high-resolution quasielastic neutron scattering
spectroscopy to study the single-particle dynamics of water
molecules on the surface of hydrated DNA samples. Both H$_{2}$O
and D$_{2}$O hydrated samples were measured. The contribution of
scattering from DNA is subtracted out by taking the difference of
the signals between the two samples. The measurement was made at a
series of temperatures from 270~K down to 185~K. The Relaxing -
Cage Model was used to analyze the quasielastic spectra. This
allowed us to extract a $Q$-independent average translational
relaxation time $\langle\tau_{T}\rangle$ of water molecules as a
function of temperature. We observe clear evidence of a
fragile-to-strong dynamic crossover (FSC) at T$_{L}$ = 222 $\pm$
2~K by plotting log$\langle\tau_{T}\rangle$ $vs$. T. The
coincidence of the dynamic transition temperature T$_{c}$ of DNA,
signaling the onset of anharmonic molecular motion, and the FSC
temperature T$_{L}$ of the hydration water suggests that the
change of mobility of the hydration water molecules across T$_{L}$
drives the dynamic transition in DNA.
\end{abstract}


\maketitle

It is known that hydrated bio-macromolecules show sharp slowing
down of their functions (kinetics of bio-chemical reactions)
within a temperature interval T $\sim$ 250-200 K
\cite{Rasmussen92}. It was also found, from neutron and X-ray
scattering, or from M\"{o}ssbauer spectroscopy, that the measured
mean-squared atomic displacement $\langle x^{2}\rangle$ of the
bio-molecules exhibits a sharp rise in the same temperature range
\cite{Rasmussen92,Ferrand93,Doster89,Tsai00,Cordone99}. This sharp
increase in $\langle x^{2}\rangle$ was taken as a sign for a
dynamic transition (or sometimes called glass-transition) in the
bio-molecules occurring within this temperature range. In most of
these papers, the authors suggest that the transition is due to a
strong rise of anharmonicity of the molecular motions above this
transition temperature \cite{Rasmussen92}. Later on, it was
demonstrated that the dynamic transition can be suppressed in dry
bio-molecules \cite{Ferrand93}, or in bio-molecules dissolved in
trehalose \cite{Cordone99}. Moreover, it can be shifted to a
higher temperature for proteins dissolved in glycerol
\cite{Tsai00}. Thus the dynamic transition can be controlled by
changing the surrounding solvent of the bio-molecules. On the
other hand, it was found some time ago from Raman scattering that
supercooled bulk water has a dynamic crossover transition at 220 K
\cite{Sokolov95}, similar to that predicted by Mode-Coupling
theory \cite{Gotze92}. Approximate coincidence of these two
characteristic temperatures, one for the slowing down of
bio-chemical activities and the sharp rise in $\langle
x^{2}\rangle$ in bio-molecules and the other for the dynamic
crossover in water, suggests a relation between the dynamic
transition of bio-molecules and that of their hydration
water~\cite{Sokolov01}.

Another striking experimental fact is that this dynamic transition
temperature, as revealed by change of slope in $\langle x^{2}\rangle$
$vs.$ temperature plot, occurs at a universal temperature range from
250 to 200 K in all bio-molecules examined so far. This list includes
globular proteins, DNAs, and t-RNAs. This feature points to the plausibility
that the dynamical transitions are not the intrinsic properties of
the bio-molecules themselves but are imposed by the hydration water
on their surfaces.

However, $\langle x^{2}\rangle$ (mostly coming from hydrogen atoms)
is an integrated quantity of motion, arising from different types
of molecular motions: both vibrations and librations of hydrogen atoms
with respect to their binding center in the molecules, as well as
large amplitude transitions between conformational substates of the
macromolecule. Therefore, it is difficult to identify the microscopic
processes underlying this transition and to pin-point the actual dynamical
transition temperature from the inspection of $\langle x^{2}\rangle$
only. On the other hand, dynamical quantities, such as the self-diffusion
coefficient, the viscosity, and the structural relaxation time (or
the so-called $\alpha$-relaxation time), could show a sharper transition
as a function of temperature and pressure if there is a genuine dynamic
transition in the hydration water.

In this paper, we demonstrate decisively using high-resolution
quasielastic neutron scattering (QENS) spectroscopy that there is
a sharp dynamic crossover, identified to be a fragile-to-strong
dynamic crossover (FSC), temperature of the hydration water in DNA
at T$_{L}=222 \pm 2$ K. This change of mobility of the water
molecules across T$_{L}$ drives the dynamic transition in DNA
which happens at the same temperature. We have recently found the
same dynamic crossover temperature of T$_{L}=220$ K for hydration
water in protein lysozyme~\cite{ChenPNAS}, which further supports
our conjecture that it is a change of mobility of the hydration
water which triggers the dynamic transition in bio-molecules.

Highly polymerized (calf thymus) DNA, sodium salt, was obtained from
Sigma (D1501, batch number 091k7030) and used without further purification.
The sample was extensively lyophilized to remove any water left. The
dry DNA fibres were then hydrated isopiestically at 5$^{\circ}$C
by exposing them to water vapor in equilibrium with a NaClO$_{3}$
saturated water solution placed in a closed chamber (relative humidity,
RH =75\%). The final hydration level was determined by thermo-gravimetric
analysis and also confirmed by directly measuring the weight of absorbed
water. This hydration level corresponding to about 15 water molecules
per base pairs was chosen to have the primary hydration sites almost
completely filled (i.e. one monolayer of water). This latter condition
corresponds to equilibration against RH=80\%~\cite{Falk62} and about
20 water molecules per base pairs~\cite{Auffinger00}. A second sample
was then prepared using D$_{2}$O in order to subtract out the incoherent
signal from the DNA hydrogen atoms. Both hydrated samples had the
same water or heavy water/dry DNA molar ratio. Differential scanning
calorimetry analysis was performed in order to detect the absence
of any feature that could be associated with the presence of bulk-like
water.

High-resolution incoherent QENS spectroscopy method is used to
determine the temperature dependence of the average translational
relaxation time $\left\langle \tau_{T}\right\rangle $ for the
hydration water. Because neutrons are predominantly scattered by
an incoherent process from the hydrogen atoms in water,
high-resolution QENS technique is an appropriate tool for the
study of diffusional process of water molecules. Using the
High-Flux Backscattering Spectrometer (HFBS) in NIST Center for
Neutron Research (NCNR), we were able to measure the $Q$-dependent
relaxation time $\tau_{T}(Q)$ (in Eq. 1) from $\approx400$ $ps$ to
$\approx5$ $ns$ over the temperature range of 270 K to 185~K,
spanning both below and above the FSC temperature. For the chosen
experimental setup, the spectrometer has an energy resolution of
0.8~$\mu$eV and a dynamic range of $\pm$ 11~$\mu$eV~\cite{Meyer},
in order to be able to extract the broad range of relaxation times
covering both the fragile and the strong regimes of the average
relaxation times $\left\langle \tau_{T}\right\rangle $ from
measured spectra.

\noindent QENS experiments measure the Fourier transform of the Intermediate
Scattering Function (ISF) of the hydrogen atoms, $F_{H}(Q,t)$, of
water molecules on the surface of DNA. Molecular Dynamics (MD) simulations
have shown that the ISF of both bulk~\cite{Gallo1} and confined~\cite{Gallo2}
supercooled water can be accurately described as a two-step relaxation:
a short-time Gaussian-like (in-cage vibrational) relaxation followed
by a plateau and then a long-time (time $>$ 1.0~$ps$) stretched
exponential relaxation of the cage. The so-called Relaxing Cage Model
(RCM)~\cite{ChenPRE}, which we use for data analysis, models closely
this two-step relaxation and has been tested extensively against bulk
and confined supercooled water through MD and experimental 
data~\cite{Gallo1,Gallo2,ChenPRE}.
By considering only the spectra with wave vector transfer $Q<1.1$~\AA $^{-1}$,
we can safely neglect the contribution from the rotational motion
of water molecule~\cite{ChenPRE}. The RCM describes the translational
dynamics of water at supercooled temperature in terms of the product
of two functions:
\begin{eqnarray}
F_{H}\left(Q,t\right) & \approx & 
F_{T}\left(Q,t\right)=F^{S}\left(Q,t\right)exp\left[-\left(t/\tau_{T}(Q)\right)^{\beta}\right],\nonumber 
\\
\tau_{T}\left(Q\right) & = &
\tau_{0}\left(0.5Q\right)^{-\gamma},\left\langle
\tau_{T}\right\rangle =\tau_{0}\Gamma\left(1/\beta\right)/\beta,
\label{decoupling}
\end{eqnarray}
\noindent where the first factor, $F^{S}\left(Q,t\right)$,
represents the short-time vibrational dynamics of the water
molecule in the cage. This function is fairly insensitive to
temperature variation, and thus can be calculated from MD
simulation. The second factor, the $\alpha$-relaxation term,
contains the stretch exponent $\beta$, and the $Q$-dependent
translational relaxation time $\tau_{T}\left(Q\right)$, which
strongly depends on temperature. The latter quantity is further
specified by two phenomenological parameters $\tau_{0}$ and
$\gamma$, the exponent controlling the power-law $Q$-dependence of
$\tau_{T}\left(Q\right)$. $\left\langle \tau_{T}\right\rangle $ is
a $Q$-independent quantity where $\Gamma$ is the gamma function.
It essentially gives a measure of the structural relaxation time
of the hydrogen-bond cage surrounding a typical water molecule.
The temperature dependence of the translational relaxation time is
then calculated from three fitted parameters, $\tau_{0}$, $\beta$,
and $\gamma$, by analyzing a group of nine quasi-elastic peaks at
different $Q$ values simultaneously.

\begin{figure}
\begin{center}
\includegraphics[width=8.6 cm]{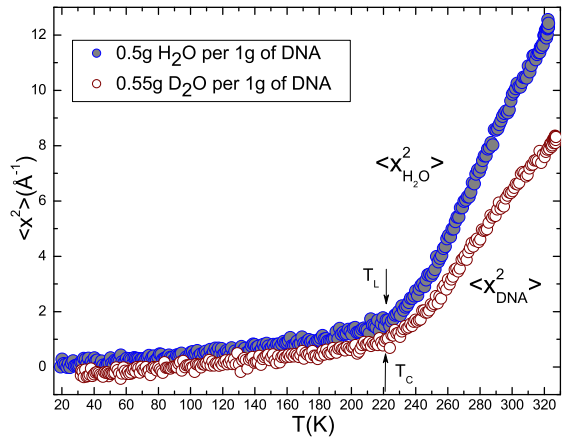}
\end{center}
\caption{Mean-squared atomic displacement $\langle x^{2}\rangle$
of all the hydrogen atoms extracted from the Debye-waller factor
measured by an elastic scan, as a function of temperature for
H$_{2}$O hydrated and D$_{2}$O hydrated DNA samples. The solid
circles represent $\langle x^{2}\rangle$ dominated by
contributions from H-atoms in hydration water, while the empty
circles are those dominated by H-atoms contained in DNA molecules.
One can clearly see that both curves have a sharp transition of
slope around 220~K indicating that the dynamic crossover
temperatures of the DNA (T$_{C}$) and the hydration water
(T$_{L}$) are approximately the same.}
\label{fig1}
\end{figure}

Fig.~\ref{fig1} shows the mean-squared hydrogen atom displacements
obtained by a method of elastic scan for hydrogen atoms in
hydration water ($\langle x_{H_{2}O}^{2}\rangle$ ) and in DNA
molecules ($\langle x_{DNA}^{2}\rangle$), respectively. One sees
that at low temperatures up to their respective crossover
temperatures, both curves have a gentle linear temperature
dependence. But above the crossover temperatures, they both rise
sharply with different slopes. We call the crossover temperature
of the former T$_{L}$, and that of the latter T$_{C}$, both have
values approximately 220~K. This shows that the dynamic crossover
phenomenon of DNA and its hydration water is highly correlated,
and occurs at the same temperature. As we shall see, this
temperature can be defined much better for the hydration water in
a dynamic measurement.

\begin{figure}
\begin{center}
\includegraphics[width=8.6 cm]{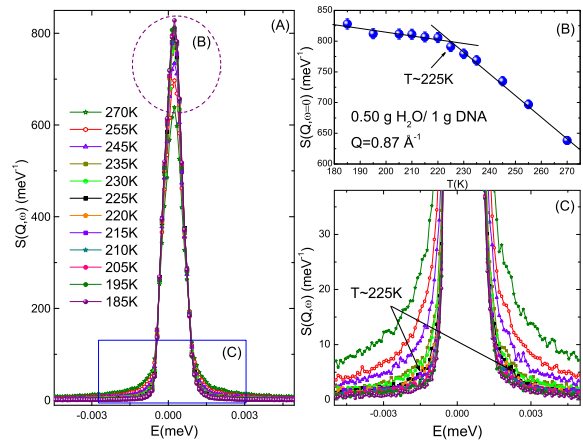}
\end{center}
\caption{Measured neutron spectra. Panel A shows normalized QENS
spectra at $Q=$~0.87 \AA$^{-1}$ at a series of temperatures.
Panels (B) and (C) display respectively the heights of the peak
(B) and the wings of the peak (C), at those temperatures. One
notes from panel (B) a cusp-like transition signaling the rate of
change of peak height from a steep high temperature region to a
slower low temperature region at a crossover temperature of about
225~K. The error bars are of the size of the data points. Panel
(C) indicates a similar change of the rate of increase of the
width at a similar crossover temperature. In this panel, the
scatter of the experimental points gives an idea of the error
bars.}
\label{fig2}
\end{figure}

We show in Fig.~\ref{fig2}, as an example, a complete set
(temperature series) of QENS area-normalized spectra. The
broadening of the quasi-elastic peaks at the wing becomes more and
more noticeable as temperature increases. At the same time, the
peak height decreases accordingly because the area is normalized
to unity. In panel B, we plot the peak height as a function of
temperature. It is noticeable that the rate of increase as a
function of temperature is different across the temperature 225 K.
 From panel C, we may notice, from the wings of these spectral
lines, that two groups of curves, 270-230~K and 220-185~K, are
separated by the curve at a temperature 220~K. This visual
information, obtained from the spectra before data analysis,
reinforces the results of the detailed line shape analysis to be
shown later in Fig.~\ref{fig4}, that there is an abrupt dynamical
transition at T$_{L}$ = 220~K.

\begin{figure}
\begin{center}
\includegraphics[width=8.6 cm]{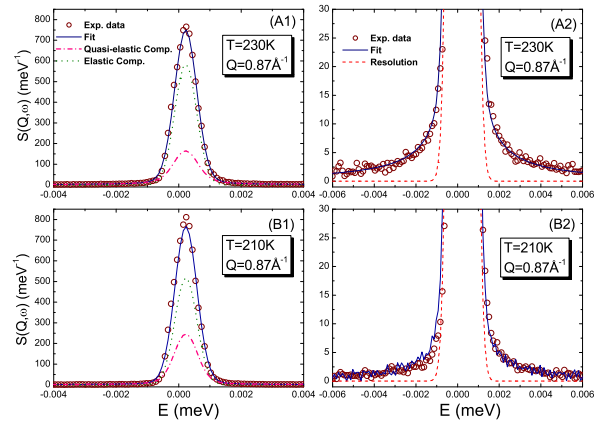}
\end{center}
\caption{RCM analyses of the QENS spectra at two temperatures,
above and below the crossover temperature. Panels (A1) and (B1)
show the results of the full analyses of the spectra at two
temperatures. Panels (A2) and (B2) indicate the detail fittings of
the wings. The resolution function is also indicated in the
figure. One can see a sharpening of the quasi-elastic peak as
temperature goes below the crossover temperature T$_{L}$ at 220~K.
The scatter of the experimental points gives an idea of the error
bars.}
\label{fig3}
\end{figure}

\begin{figure}
\begin{center}
\includegraphics[width=7 cm]{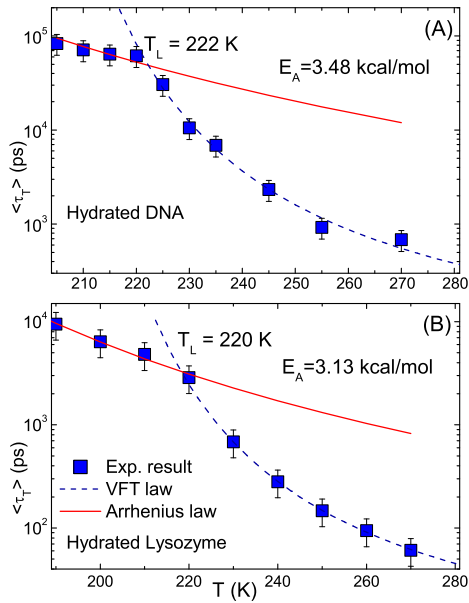}
\end{center}
\caption{The extracted $Q$-independent average translational
relaxation time $\langle\tau_{T}\rangle$ from fitting of the
quasielastic spectra plotted in log scale against temperature.
Panel (A) is a result from hydrated DNA, whereas Panel (B) is the
same quantity measured in hydrated lysozyme shown for
comparison~\cite{ChenPNAS}. There is a clear evidence in both
cases a well-defined cusp-like dynamic crossover behavior
occurring at T$_{L}$ indicated in the respective figures. The
dashed lines represent fitted curves using VFT law, while the
solid lines the fitting according to Arrhenius law. T$_{L}$ in
both cases occurs at $222 \pm 2$~K.} \label{fig4}
\end{figure}

Fig.~\ref{fig3} shows the result of RCM analyses of the spectra
taken at Q = 0.87 \AA $^{-1}$ for temperatures 230 K (panel A) and
210 K (panel B), before and after the T$_{L}$, respectively. In this
figure, we display the instrument resolution function purposely for
comparison with the measured spectrum. RCM, as one can see, reproduces
the experimental spectral line shapes of hydration water quite well.
The broadening of the experimental data over the resolution function
leaves enough dynamic information to be extracted by RCM.

In Fig.~\ref{fig4}, we present the temperature dependence of the
average translational relaxation time, $\left\langle
\tau_{T}\right\rangle $, for the hydrogen atom in a water molecule
calculated by Eq. 1. It is seen that, in the temperature range
from 270 to 230 K, $\left\langle \tau_{T}\right\rangle$ obeys
Vogel-Fulcher-Tammann (VFT) law, a signature of fragile liquid,
quite closely. But at T = 222 K it suddenly switches to an
Arrhenius law, a signature of a strong liquid. So we have a clear
evidence of FSC in a cusp form. The T$_{0}$ for the fragile liquid
turns out to be 180 K, and the activation energy for the strong
liquid, E$_{A}$ = 3.48 kcal/mol. As a comparison, we plot the same
quantity in panel B for hydration water in lysozyme
protein~\cite{ChenPNAS}. It is to be noted that the crossover
temperature is sharply defined at T$_{L} = 220$~K, slightly lower
than in the DNA case.

\begin{figure}
\begin{center}
\includegraphics[width=7 cm]{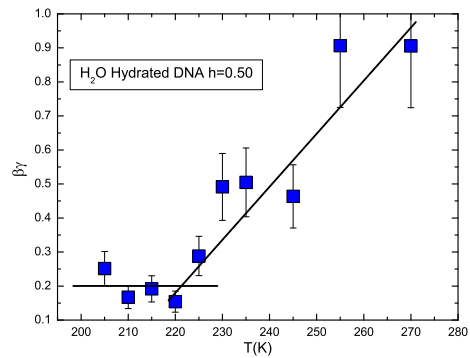}
\end{center}
\caption{Temperature dependence of the exponent $\beta\gamma$ of
the power law $Q$-dependence of ISF. }
\label{fig5}
\end{figure}

In summary, we present unequivocal evidence that there is a fragile-to-strong
dynamic crossover phenomenon observable in both DNA and protein hydration
water. Above the crossover temperature, the hydration water is more
fluid, implying having locally predominantly high-density water 
structure~\cite{Xu05,ChenPRL05},
with a not fully developed hydrogen-bond network; and below the crossover
temperature, it evolves into locally predominantly low-density water
structure, corresponding to an extensive hydrogen-bond network, which
is less-fluid. This mobility change across T$_{L}$ can be seen from
Fig.~\ref{fig5}, which shows the power law $Q$-dependence of ISF,
$\beta\gamma$, as a function of temperature (calculated by Eq. 1).
One sees from the figure that $\beta\gamma$ decreases steadily as
the temperature decreases, reaching the lowest value 0.2 at the crossover
temperature 220 K. It should be noted that for a freely diffusing
water molecule, $\beta\gamma=2$, therefore the very low value of
$\beta\gamma=0.2$ signifies a restricted mobility of the hydration
water. There is a strong evidence from MD simulations of protein hydration
water that this drastic change of mobility across the FSC triggers
the so-called glass transition in protein 
molecules~\cite{Tournier03,Tarek02,Kumar06}.
This paper supplies an experimental evidence which reinforces this
interpretation for the case of protein and DNA~\cite{Kumar06}.

The research at MIT is supported by DOE Grants DE-FG02-90ER45429 and
2113-MIT-DOE-591. EF and PB acknowledge CSGI (Florence, Italy) for
partial financial support. This work utilized facilities supported
in part by the National Science Foundation under Agreement No.DMR-0086210.
Technical support in measurements from V. Garcia-Sakai at NIST NCNR
is greatly appreciated. We benefited from affiliation with EU-Marie-Curie
Research and Training Network on Arrested Matter.

\end{document}